*Review*

# Recent Advances in Nanostructured Thermoelectric Half-Heusler Compounds

**Wenjie Xie [1],\*, Anke Weidenkaff [1], Xinfeng Tang [2], Qingjie Zhang [2], Joseph Poon [3] and Terry M. Tritt [4],\***

[1] Empa, Swiss Federal Laboratories for Materials Science and Technology, Solid State Chemistry and Catalysis, CH-8600 Dubendorf, Switzerland; E-Mail: anke.weidenkaff@empa.ch

[2] State Key Laboratory of Advanced Technology for Materials Synthesis and Processing, Wuhan University of Technology, Wuhan 430070, China; E-Mails: tangxf@whut.edu.cn (X.T.), zhangqj@whut.edu.cn (Q.Z.)

[3] Department of Physics, University of Virginia, Charlottesville, VA 22904-4714, USA; E-Mail: sjp9x@virginia.edu

[4] Department of Physics and Astronomy, Department of Materials Science & Engineering, Clemson University, Clemson, SC 29634-0978, USA

\* Author to whom correspondence should be addressed; E-Mails: wenjie.xie@empa.ch (W.X.); ttritt@clemson.edu (T.M.T.); Tel.:+41-58-765-4861 (W.X.); +1-864-656-5319 (T.M.T); Fax: +41-58-765-4019 (W.X.); +1-864-656-0805 (T.M.T.).

**Abstract:** Half-Heusler (HH) alloys have attracted considerable interest as promising thermoelectric (TE) materials in the temperature range around 700 K and above, which is close to the temperature range of most industrial waste heat sources. The past few years have seen nanostructuing play an important role in significantly enhancing the TE performance of several HH alloys. In this article, we briefly review the recent progress and advances in these HH nanocomposites. We begin by presenting the structure of HH alloys and the different strategies that have been utilized for improving the TE properties of HH alloys. Next, we review the details of HH nanocomposites as obtained by different techniques. Finally, the review closes by highlighting several promising strategies for further research directions in these very promising TE materials.

## 1. Introduction

The imperative demand for alternative and sustainable energies and energy conversion technologies to reduce our global reliance on fossil fuels leads to important regimes of research, including that of thermoelectricity. Thermoelectricity is the simplest technology applicable to achieve the direct conversion between heat and electricity. The dimensionless figure of merit ($ZT$) is determined by $ZT = \alpha^2\sigma T/\kappa = \alpha^2 T/\rho(\kappa_L + \kappa_e)$, where $\alpha$ is the Seebeck coefficient (*i.e.*, thermopower), $\sigma$ the electrical conductivity, $\rho$, the electrical resistivity, $\kappa$, the thermal conductivity (including the lattice thermal conductivity, $\kappa_L$, and the carrier thermal conductivity, $\kappa_e$), and $T$ the absolute temperature. A high efficiency thermoelectric (TE) device consists of legs (pellets) each made of high dimensionless figure of merit *n*-type or *p*-type material. These *n*-type or *p*-type legs are connected electrically in series and thermally in parallel to form a TE module or device. According to the definition of $ZT$, then a high $ZT$ material is essentially a "phonon-glass electron-crystal" (PGEC) system with a large Seebeck coefficient [1,2]. Apparently, it is difficult to satisfy these criteria in a material with a simple crystal structure since the three physical properties ($\alpha$, $\sigma$, and $\kappa$) that govern the $ZT$ are interdependent: a modification to any of these properties often adversely affects the other properties. Thus, the $ZT$ values of most state-of-the-art TE materials, such as $Bi_2Te_3$ [3–5], SiGe [6–8], and PbTe [9–12], are below or around to the value of unity, *i.e.*, $ZT_{max} \approx 1$. The conversion efficiency of TE device is determined by both the Carnot efficiency and the $ZT$ values of the TE materials. For example, the maximum conversion efficiency of TE power generation, $\eta$, is given by:

$$\eta = \left(\frac{T_H - T_C}{T_H}\right)\left[\frac{\sqrt{1 + ZT_m} - 1}{\sqrt{1 + ZT_m} + T_C/T_H}\right] \tag{1}$$

where $T_H$ is the hot side temperature, $T_C$ the cold side temperature, and $T_m$ the mean temperature. Of course, the first term in parenthesis is the Carnot efficiency. For a given working temperature, e.g., $T_C = 300$ K and $T_H = 800$ K, the maximum conversion efficiency dependence of $ZT$ values can be obtained by Equation 1 and illustrated in Figure 1. As shown in Figure 1, for $T_C = 300$ K and $T_H = 800$ K, if the $ZT = 1$, the maximum $\eta$ is around 10%, which is much lower than the Carnot efficiency (62.5%). Although the low efficiency of current TE materials is the main drawback of TE technology, nevertheless TE technology still plays an important role in the recovery of industrial or automotive waste heat [13–21], utilization of solar thermal energy (infrared) [22–29], and the application of solid-state refrigeration [30–33]. It is important to note and remember that waste heat is a 0% efficiency process and thus any cost effective technology that can recover some part of this waste heat is very important. Furthermore, unused thermal energy is very abundant, such as industrial or automotive waste heat.

The key goals of TE research are to discover brand new classes of materials with high TE performance and/or improve the performance of the existing well-known materials, such as $Bi_2Te_3$, SiGe, and PbTe alloys. These materials ideally, should also be stable at high temperature, possess low contact resistance and exhibit low parasitic losses. Among the wide variety of TE materials, half-Heusler (HH) compounds are intermetallic compounds formulated as MNiSn [34–48] and MCoSb [49–62] (M = Ti, Zr, or Hf) and it was pointed out as one kind of new promising TE materials in the temperature range around 700 K and above. Band structure calculations [63–66] show that many

HH systems with 18-valence electrons are narrow band gap semiconductors, which will result in a high effective mass and a large Seebeck coefficient. For instance, the Seebeck coefficient of HH compounds TiNiSn, ZrNiSn and HfNiSn are in the range of $-200$ to $-400$ $\mu$V/K (*n*-type) at room temperature [36,62], which is higher than that of the state-of-the-art $Bi_2Te_3$ compounds. For this reason, HH compounds have attracted considerable attention as a promising candidate material for moderate and high temperature TE power generation applications. However, the biggest disadvantage for the HH compounds is their relatively high thermal conductivity. Two of the authors (Poon and Tritt) have been collaboratively working on these materials for over more than a decade. For the most widely investigated MNiSn and MCoSb (M = Ti, Zr, or Hf) systems, the thermal conductivity is on the order of 10 $Wm^{-1}K^{-1}$ at room temperature [34,36,67,68], which is almost 5 times higher than that of commercial $Bi_2Te_3$ compounds [3–5]. Therefore, over last two decades, the main research efforts have focused extensively on methods and techniques in order to significantly decrease the thermal conductivity of these HH compounds, while simultaneously maintaining their high power factors, *PF* ($PF = \alpha^2\sigma = \alpha^2/\rho$). The power factor reflects primarily the electronic transport component of the figure of merit, *ZT*.

**Figure 1.** The maximum conversion efficiency dependence of figure of merit (*ZT*) values under certain working temperature.

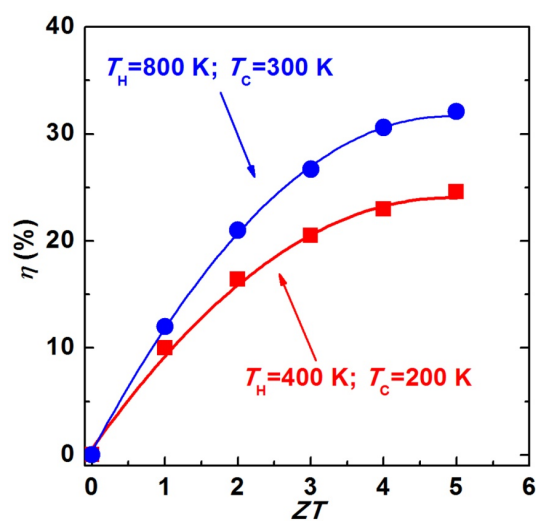

Over the past several years, one successful strategy that has been used to decrease the thermal conductivity in several TE materials is that of nanostructuring or introducing nano-scale secondary phases into the host matrix to form nanocomposites. Indeed, the nanostructuring approach significantly enhances the *ZT* values of HH nanocomposites by significantly reducing the lattice thermal conductivity [69–78]. However, more recently, it is reported that, in addition to the reduction in the lattice thermal conductivity, nanostructuring can also increase the power factor, *PF*, in some HH nanocomposites [79,80]. In other words, it is possible that in some material systems the three interrelated physical properties ($\alpha$, $\sigma$, and $\kappa$) can be decoupled by introducing nanostructures.

Herein, we review the recent progress in improving the TE performance of several HH compounds by utilizing the strategy of forming HH nanocomposites. Before the detailed review of HH nanocomposites, we would like to provide some important background by introducing the structure of HH compounds, and highlight some of the more typical strategies that have been previously used in order to enhance the TE properties of HH compounds. After that, we will review the recent progress in HH nanocomposites, which can be classified according to the various types of microstructures. Finally, we finish with a discussion of challenges and promising strategies for further research directions in these materials.

## 2. Structure of HH Compounds and Typical Strategies of Enhancing *ZT* for HH

The half-Heusler phase is one kind of the best known intermetallic compounds, and it is represented by the general formula ABC, where A and B are transition metals, and C is an *sp* metalloid or metal [62,81]. The crystal structure of a typical HH compound ABC is MgAgAs type with space group of $F\bar{4}3m$, and is shown in Figure 2. In the structure shown in Figure 2, it consists of three filled interpenetrating face-centered cubic (fcc) sublattices and one vacant fcc sublattice. One should note that the full Heusler material would contain all four interpenetrating sublattices and would be of the form, $AB_2C$. In the HH compounds, the elements A and C form a rock salt structure and B is located at one of the two body diagonal positions (1/4, 1/4, 1/4) in the cell with leaving the other one (3/4, 3/4, 3/4) unoccupied. The more detail introduction of HH structure can be found in review by Graf *et al.* [82].

**Figure 2.** The crystal structure of a typical Half-Heusler (HH) compound ABC.

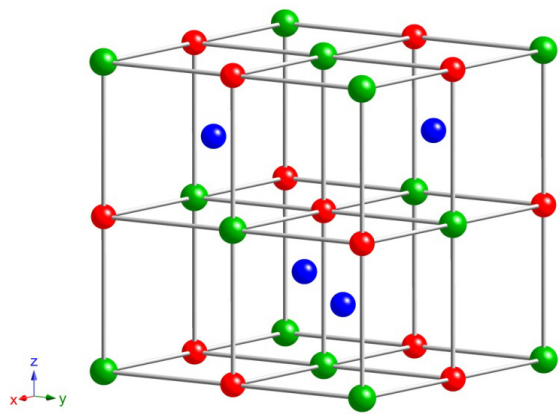

One of the first strategies for enhancing *ZT* values (not only for HH materials, but for all TE materials) is by tuning the carrier concentration via ***doping***. Because the electrical conductivity σ increases with an increase in carrier concentration, and while the Seebeck coefficient α decreases, the power factors ($PF = \alpha^2\sigma$) typically exhibit their highest values at carrier concentrations between $10^{19}$ and $10^{21}$ carriers per $cm^3$ [83]. This indicates that heavily doped semiconductors, either narrow band gap or degenerate semiconductors, usually result in the best TE materials. For HH compounds, due to their specific structural features, it is possible to dope each of the three occupied fcc sublattices individually for tuning and optimizing the carrier concentration. Moreover, significant doping (or atomic substitutions) in the HH compounds can also introduce point defects, mass fluctuation scattering and strain field effects that can be utilized in order to effectively scatter short- and mid-wavelength heat-carrying phonons, which will significantly reduce the lattice thermal conductivity. In the most widely investigated MNiSn and MCoSb (M = Ti, Zr, or Hf) systems, it is typical to alter the number of charge carriers by doping on Sn and Sb site, and simultaneously introduce disorder by isoelectronic alloying on M site with Ti, Zr and Hf. For example, in the TiNiSn system, using $x = 1\%$–$5\%$ Sb for Sn, in TiNiSn$_{1-x}$Sb$_x$) yield the best changes in the band structure in order to effectively reduce the resistivity. Furthermore, using isoelectronic doping on the A and B sites, e.g., Ti$_{1-y}$MyNi$_{-z}$R$_z$Sn$_{1-x}$Sb$_x$ (M = Zr, or Hf) for the A site and (R = Pd or Pt) for the B site work best for manipulating or significantly lowering the lattice thermal conductivity. One can see that the compositions can get extremely complex. Thus, the *ZT* values of MNiSn and MCoSb (M = Ti, Zr, or Hf) systems can be significantly improved. Figure 3 presents an overview on the most promising doped *n* and *p*-type HH compounds with relatively high *ZT* values. For the doped *n*-type TiNiSn and TiCoSb systems, the reproducible highest *ZT* values are achieved to values of $ZT \approx 1$ [84,85] and $ZT \approx 0.7$ [49], respectively, which are much higher than that of the undoped compounds. One must also make note that one of the *very important attributes* of these materials is that they exhibit excellent thermal stability to temperatures as high as 900 °C in the Ti$_{1-y}$MyNi$_{-z}$R$_z$Sn$_{1-x}$Sb$_x$ system. Also by doping the *p*-type HH compounds they obtain their highest *ZT* values of $ZT \approx 0.5$ at 1000 K [86].

**Figure 3.** Overview of the most promising doped *n* and *p*-type doped HH compounds:
(**a**) *n*-type, (**b**) *p*-type. (1) ErNi$_{1-x}$Pd$_x$Sb [87]; (2) ZrNi$_{0.8}$Ir$_{0.2}$Sn [88]; (3) TiCo$_{0.85}$Fe$_{0.15}$Sb [51];
(4) ZrCoSn$_{0.1}$Sb$_{0.9}$ [89]; (5) Zr$_{0.5}$Hf$_{0.5}$CoSb$_{0.8}$Sn$_{0.2}$ [86]; (6) Ti$_{0.5}$Zr$_{0.25}$Hf$_{0.25}$Co$_{0.95}$Ni$_{0.05}$Sb [50];
(7)    Ti$_{0.6}$Hf$_{0.4}$Co$_{0.87}$Ni$_{0.13}$Sb    [49];    (8)    Zr$_{0.75}$Hf$_{0.25}$NiSn$_{0.975}$Sb$_{0.025}$    [90];
(9) Zr$_{0.4}$Hf$_{0.6}$NiSn$_{0.98}$Sb$_{0.02}$ [84]; (10) Zr$_{0.25}$Hf$_{0.25}$Ti$_{0.5}$NiSn$_{0.998}$Sb$_{0.002}$ [91].

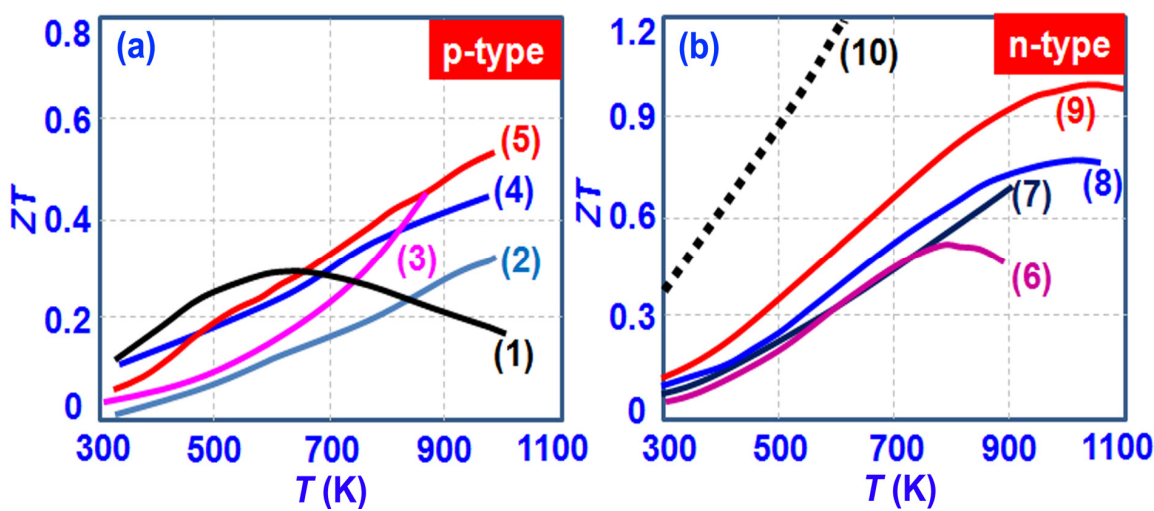

Further significant improvements in the TE properties of HH compounds are thus dependent upon new materials design approaches other than just simple doping or substitutions. The most probable of these is the recent successful strategy of utilizing a ***nanostructuring*** approach in these materials. *Nanostructuring* has become a new paradigm approach of TE materials research that has been widely used to improve the *ZT* of several bulk TE materials over the past several years [92-100]. In the nanostructured bulk materials, or nanocomposites, the presence of multiple length-scales down to a few nanometers ($\ell_{NC}$), provides new avenues for phonon scattering in addition to the mass fluctuation alloying ($\ell_{MF}$), grain boundary or surface scattering ($\ell_B$), phonon-phonon interactions ($\ell_{ph-ph}$), and strain fields ($\ell_{SF}$), which all can occur in parallel and thus each adds to the process according to the Matthiessen's rule, given below in Equation 2, with the shortest scattering process dominating.

$$\frac{1}{\ell} = \frac{1}{\ell_{NC}} + \frac{1}{\ell_{MF}} + \frac{1}{\ell_B} + \frac{1}{\ell_{ph-ph}} + \frac{1}{\ell_{SF}} \qquad (2)$$

Therefore, all these various scattering processes cover a wide range of phonon wavelengths in order to more effectively reduce the lattice thermal conductivity. Meanwhile, it is also expected that, in a system with a semiconductor host matrix with metallic nanoinclusions, the Seebeck coefficient can be increased without significantly sacrificing electrical conductivity by interface charge carrier energy filtering effect [101–105]. Many different HH nanocomposites were investigated in the past with regards to improving their TE properties, and the details will be reviewed in **section 3**. Implementation of the nanostructuring process in HH compounds has indeed led to enhancement of *ZT*, which is due to the significant reduction of the lattice thermal conductivity in most cases [69–78], and also a substantial improvement of the power factor [79,80].

### 3. Nanostructuring Enhances *ZT* of HH Nanocomposites

In this section, we categorize the HH nanocomposites in terms of the type of microstructure: (1) micro-scale HH matrix with nano-scale inclusions (shown in Figure 4a), and (2) nano-scale HH phase with nanoinclusions (shown in Figure 4b). For the two kinds of nanocomposites, there are also several different preparation methods for each of them. For example, nanoinclusions in a micro-scale HH matrix can be prepared by in-situ formation [79,80,106] as well as *ex-situ* mixing [71,77,78,107–110], and the nano-scale HH phase can be obtained by either ball milling [53,67,72,73,111] or melt spinning [74,112,113]. It is not our intent (which would be quite daunting) to cover all the possible preparation methods for HH nanocomposites in this article. Instead, the purpose of this review lies in depicting the most representative HH nanocomposites with particular emphasis on the relation between their nanostructure and the resulting TE properties from results obtained over the last several years.

**Figure 4.** Two types of HH nanocomposites: (**a**) micro-scale HH matrix with nano-scale inclusions; (**b**) nano-scale HH phase with nanoinclusions.

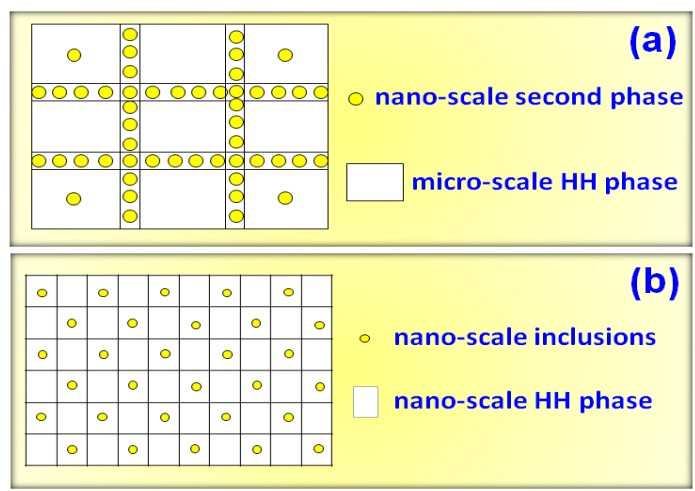

*3.1. Micro-Scale HH Matrix with Nanoinclusions*

Nanoinclusions can be introduced by either mechanical mixing (*ex-situ*) or nanoprecipitation (*in-situ*) processes.

3.1.1. *Ex-Situ* Approach-Mechanical Mixing

It is relatively straightforward and uncomplicated to obtain nanocomposites with nanoinclusions by mechanical mixing. Usually some commercial nanoparticles (or other nanomaterials) with very uniform size distributions are available, and one can then purchase the uniform nanoparticles and subsequently mix the nanoparticles with HH matrix compound by ball milling in order to form nanocomposites. $\gamma$-$Al_2O_3$ [78], $ZrO_2$ [77,108,110], $C_{60}$ [109], $WO_3$ [107], and NiO [71] nanoscale secondary phases are chosen as nanoinclusions in the HH matrix, and the effects of the nanoinclusions on the electrical and thermal transport properties of HH compounds have been systemically investigated.

Huang *et al.* prepared ZrNiSn-$ZrO_2$ nanocomposites with ~20 nm $ZrO_2$ by mechanical mixing followed by spark plasma sintering, and investigated the effects of $ZrO_2$ nanoinclusions on the TE transport properties of ZrNiSn-$ZrO_2$ nanocomposites [77]. The microstructures of un-sintered $ZrO_2$ nanoparticles and spark plasma sintered (SPS) ZrNiSn/$ZrO_2$ nanocomposites are presented in Figure 5. As shown in Figure 5a, the size of $ZrO_2$ nanoparticles is very small (~20 nm) and uniform. After mixed and sintered, scanning electron microscopy (SEM) investigation, shown in Figure 5b, indicates that 20 nm $ZrO_2$ nanoparticles do not grow very much at all during the sintering process. In addition, the electron probe micro-analysis (EPMA) observation shows that the $ZrO_2$ nanoparticles are mainly distributed at the grain boundaries. Such small size $ZrO_2$ nanoparticles as nanoinclusions were chosen with the clear purpose to scatter heat-carrying phonons, and subsequently reduce the resulting lattice thermal conductivity. By introducing $ZrO_2$ nanoinclusions, the thermal conductivity of ZrNiSn- 6 vol% $ZrO_2$ indeed reduces by 35% as compared with that of ZrNiSn without nanoinclusions (Figure 6c). It is reasonable that the $ZrO_2$ nanoinclusions, mainly dispersed along the grain boundaries of the materials, could act as phonon scattering centers in order to significantly scatter the heat-carrying phonons. Thus, the phonon mean-free-path should be considerably reduced, and the lattice thermal conductivity was significantly reduced. However, it is not unexpected that resistivity of ZrNiSn- 6 vol% $ZrO_2$ also increases by ~30% compared with that of ZrNiSn without nanoinclusions. Although Huang *et al.* did not give a convincing explanation for the increase in resistivity at that time, we believe that such small $ZrO_2$ nanoinclusions with the size of 20 nm embedded in ZrNiSn matrix not only significantly scatter the phonons, but they also consequently result in scattering the electrons. The highest *ZT* of ZrNiSn- 6 vol% $ZrO_2$ nanocomposite achieves a value of $ZT \approx 0.25$ at 900 K, which is about a 15% improvement as compared with that of pure ZrNiSn compounds.

**Figure 5.** (**a**) TEM image of un-sintered $ZrO_2$ nanoparticles; (**b**) SEM image of spark plasma sintered ZrNiSn/2 vol% $ZrO_2$ nanocomposite. Reproduced with permission from Reference [77]. Copyright 2004, Elsevier.

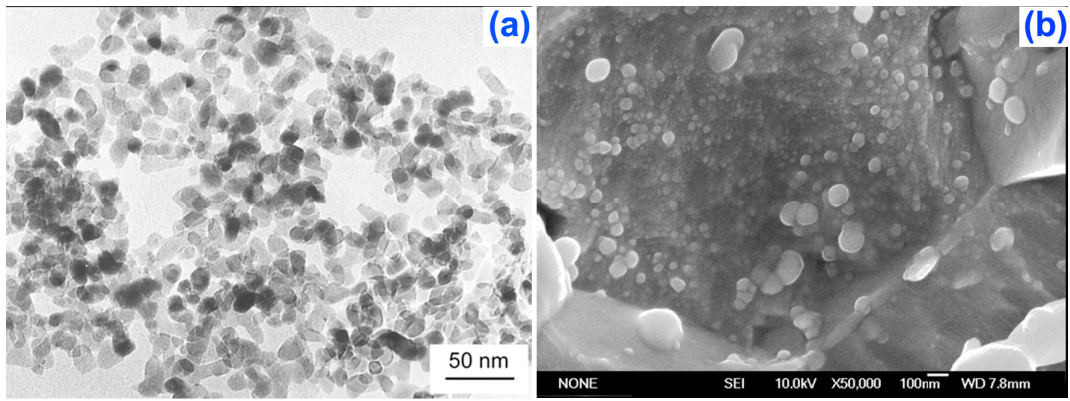

**Figure 6.** Thermoelectric (TE) transport properties of ZrNiSn-$ZrO_2$ nanocomposites: (**a**) resistivity; (**b**) the Seebeck coefficient; (**c**) thermal conductivity; and (**d**) $ZT$ values. The data is adopted from Reference [77].

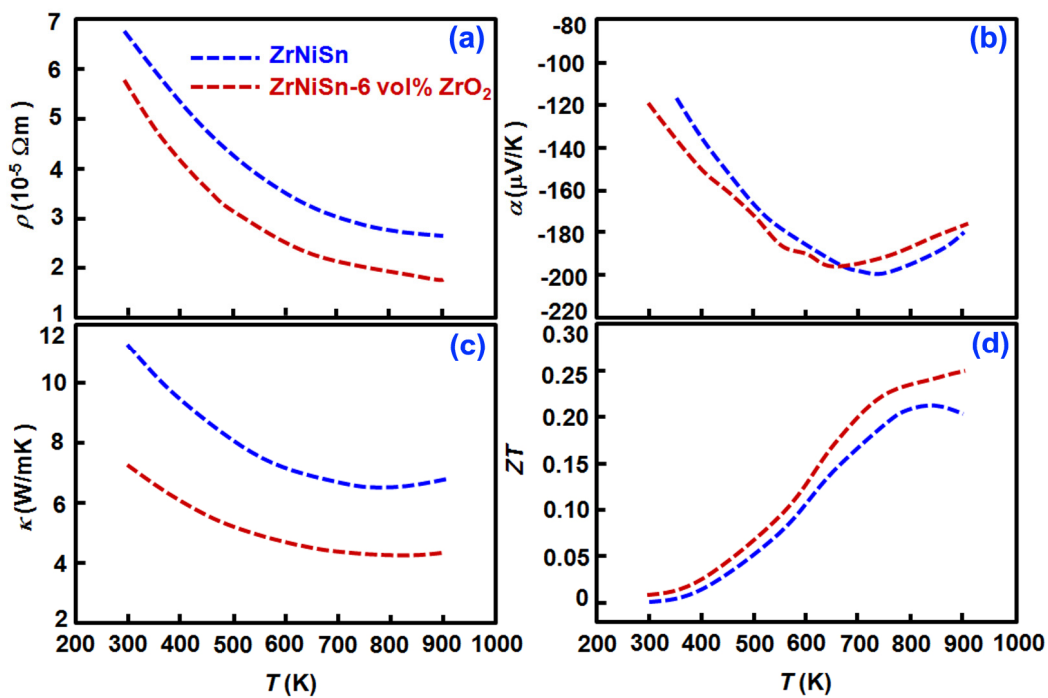

Subsequently, the same procedure has been adopted to prepare the *n*-type $Zr_{0.5}Hf_{0.5}Ni_{0.8}Pd_{0.2}Sn_{0.99}Sb_{0.01}$-$ZrO_2$ nanocomposites [108]. The microstructure and distribution of $ZrO_2$ nanoinclusions in $Zr_{0.5}Hf_{0.5}Ni_{0.8}Pd_{0.2}Sn_{0.99}Sb_{0.01}$ matrix are almost the same as that of in previous ZrNiSn-$ZrO_2$ shown in Figure 5. The electrical and thermal transport properties of $Zr_{0.5}Hf_{0.5}Ni_{0.8}Pd_{0.2}Sn_{0.99}Sb_{0.01}$-$ZrO_2$ nanocomposites are presented in Figure 7. As shown in Figure 7c, compared with the lattice thermal conductivity of $Zr_{0.5}Hf_{0.5}Ni_{0.8}Pd_{0.2}Sn_{0.99}Sb_{0.01}$ without nanoinclusions, the reduction in these three nanocomposites with 3, 6, and 9 vol% $ZrO_2$ nanoinclusions at room temperature are approximately 5%, 14%, and 34%, respectively. It is due to the realization that the $ZrO_2$ nanoinclusions do not just act as phonon scattering centers to scatter heat carrying phonons, but they also induce defect phonon scattering in order to further decrease the lattice thermal conductivity. In addition, it is worth noting that, although the carrier concentration of $Zr_{0.5}Hf_{0.5}Ni_{0.8}Pd_{0.2}Sn_{0.99}Sb_{0.01}$-$ZrO_2$ nanocomposites increases with increasing the $ZrO_2$ volume ratio, the Seebeck coefficients of $Zr_{0.5}Hf_{0.5}Ni_{0.8}Pd_{0.2}Sn_{0.99}Sb_{0.01}$-$ZrO_2$ nanocomposites with 6 and 9 vol% $ZrO_2$ nanoinclusions show ~10% enhancement at 800 K compared with that of the $Zr_{0.5}Hf_{0.5}Ni_{0.8}Pd_{0.2}Sn_{0.99}Sb_{0.01}$ matrix. The possible explanation of the improvement in the Seebeck coefficient is that the $ZrO_2$ nanoinclusions induce a potential barrier to scatter electrons, resulting in an enhancement in the scattering parameter. Due to primarily the significant reduction in lattice thermal conductivity as well as enhancement in Seebeck coefficient, then $Zr_{0.5}Hf_{0.5}Ni_{0.8}Pd_{0.2}Sn_{0.99}Sb_{0.01}$ with 9 vol% $ZrO_2$ nanocomposites achieves the highest $ZT$ of 0.75 at 800 K, a ~30% improvement over the $Zr_{0.5}Hf_{0.5}Ni_{0.8}Pd_{0.2}Sn_{0.99}Sb_{0.01}$ matrix.

**Figure 7.** TE transport properties of $Zr_{0.5}Hf_{0.5}Ni_{0.8}Pd_{0.2}Sn_{0.99}Sb_{0.01}$-x vol% $ZrO_2$ nanocomposites: (**a**) the Seebeck coefficient; (**b**) power factor; (**c**) lattice thermal conductivity; and (**d**) calculated $ZT$ values for $Zr_{0.5}Hf_{0.5}Ni_{0.8}Pd_{0.2}Sn_{0.99}Sb_{0.01}$ without nanoinclusions and with 9 vol% $ZrO_2$. For the (a), (b) and (c), reproduced with permission from Reference [108]. Copyright 2006, American Institute of Physics.

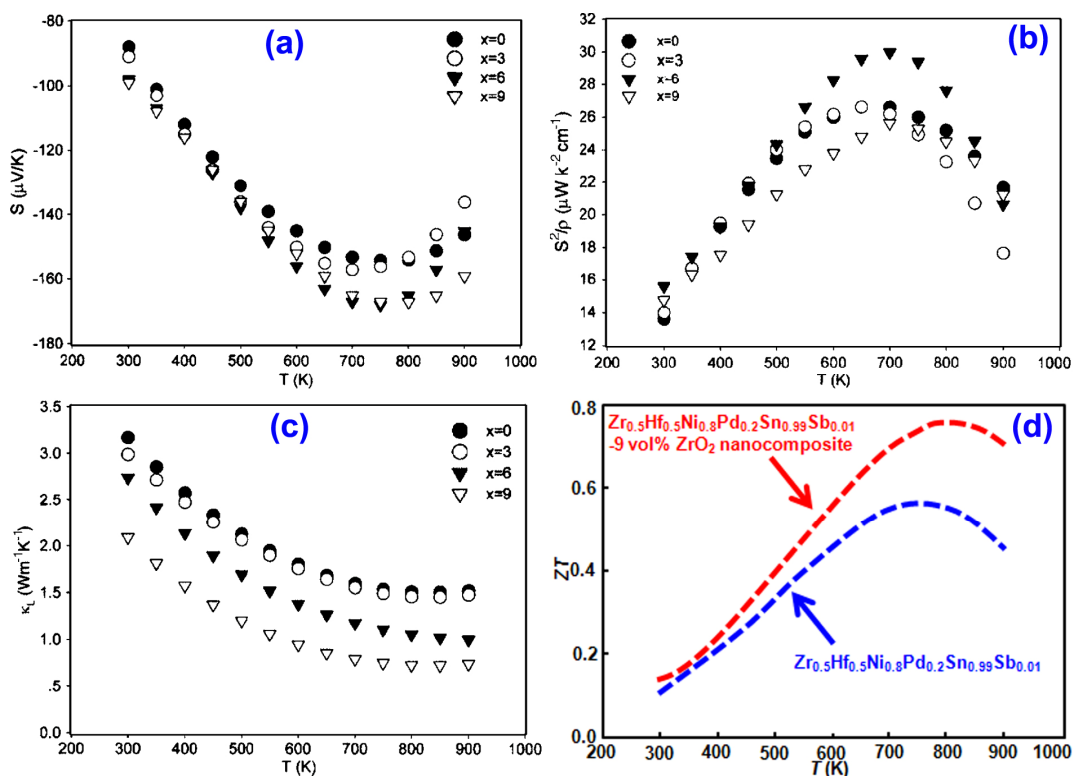

Nanoscale $ZrO_2$ particles as nanoinclusions can enhance the TE performance of *n*-type ZrNiSn based nanocomposites, so consequently, it is very reasonable to imagine as to whether one can apply the same approach to enhance the TE performance of *p*-type HH nanocomposites. Recently, Poon *et al.* investigated the relation between microstructure and TE properties of *p*-type $Hf_{0.3}Zr_{0.7}CoSn_{0.3}Sb_{0.7}$ with 20–300 nm $ZrO_2$ nanocomposites [110]. The TEM and SEM images (Figure 8) show that $ZrO_2$ assemble in the HH grain boundaries. Due to a combination of enhanced power factor and reduced electronic or lattice thermal conductivity, the highest *ZT* of Hf$_{0.3}$Zr$_{0.7}$CoSn$_{0.3}$Sb$_{0.7}$ nanocomposites, with 1 vol% of ZrO$_2$ nanoinclusions, achieves *ZT* ≈ 0.8 at 970 K (Figure 9), which increases by ~23% as compared with that of matrix ingot.

**Figure 8.** TEM (**a**) and SEM (**b**) image of Hf$_{0.3}$Zr$_{0.7}$CoSn$_{0.3}$Sb$_{0.7}$ dispersed with 2% ZrO$_2$ nanoparticles showing nanoinclusions on the grain boundary of main matrix. (**c**) and (**d**) Diffraction and energy dispersive spectroscopy (EDS) patterns from nanoparticles (location B) and main matrix (location E) in figure a, respectively. Reproduced with permission from Reference [110]. Copyright 2011, Cambridge University Press.

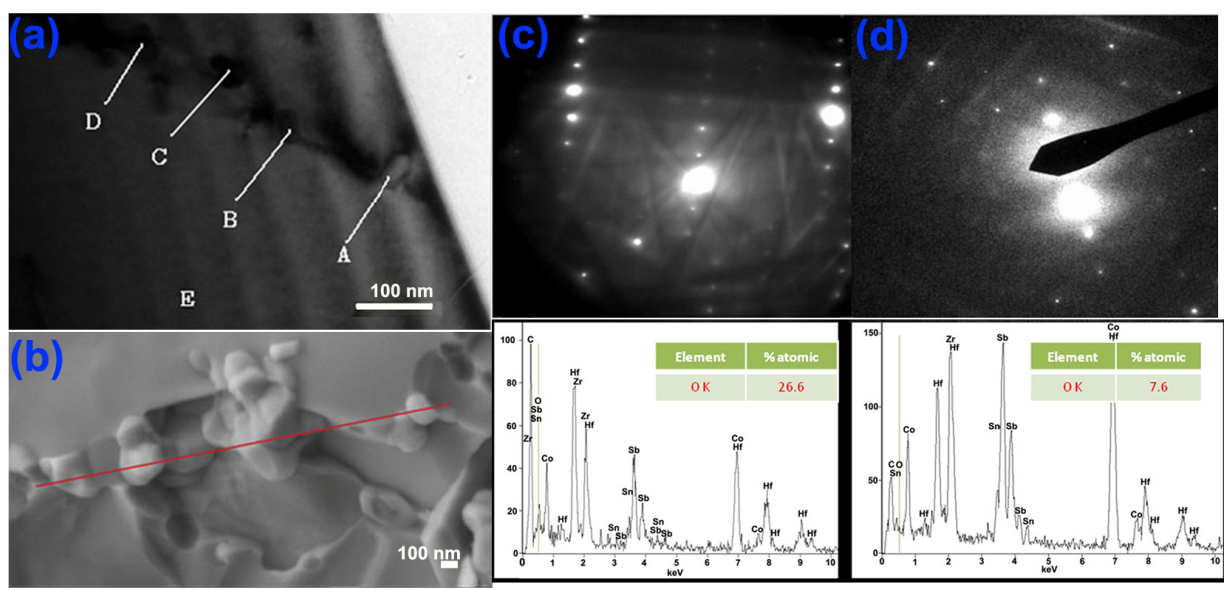

It is apparent that the main contribution to the enhancement in *ZT* of HH nanocomposites prepared by the *ex-situ* approach (mechanical mixing) originates from a significant reduction in the lattice thermal conductivity as a result of the additional phonon scattering induced by the secondary phase nanoinclusions. Although Chen *et al.* and Poon *et al.* observed the potential energy barrier scattering (or *energy filtering effect*) in HH-ZrO$_2$ nanocomposites, it is still not strong enough to significantly change the Seebeck coefficient as well as a pure or significant energy filtering process. Moreover, one of the main disadvantages of mechanical mixing is that usually it will also induce extra defects that in turn results in the additional scattering of the charge carriers, resulting in a noticeable increase in resistivity. Such an increase in resistivity can also give rise to an unexpected decrease of *ZT* for ZiNiSn-γ-Al$_2$O$_3$ [78] and Zr$_{0.5}$Hf$_{0.5}$Ni$_{0.8}$Pd$_{0.2}$Sn$_{0.99}$Sb$_{0.01}$-WO$_3$ [107] nanocomposites. In the most ideal case, the nanostructures will scatter phonons more effectively than they scatter electrons, or even increase the electrical conductivity by other physical mechanisms, such as the modulation doping effect [114,115]. To the best of our knowledge, so far, there are no reports that the three interrelated physical properties (α, σ or ρ, and κ) can be simultaneously optimized in nanocomposites when they are prepared by the *ex-situ* approach (mechanical mixing). However, such cases are observed in the situation of *in-situ* forming nanocomposites.

**Figure 9.** TE transport properties of 0% (square), 1% (triangle), and 2% (star) $ZrO_2$ nanoparticles dispersed *p*-type $Hf_{0.3}Zr_{0.7}CoSn_{0.3}Sb_{0.7}$: (**a**) the Seebeck coefficient; (**b**) electrical resistivity; (**c**) thermal conductivity; and (**d**) *ZT* values. Reproduced with permission from Ref. [110]. Copyright 2011, Cambridge University Press.

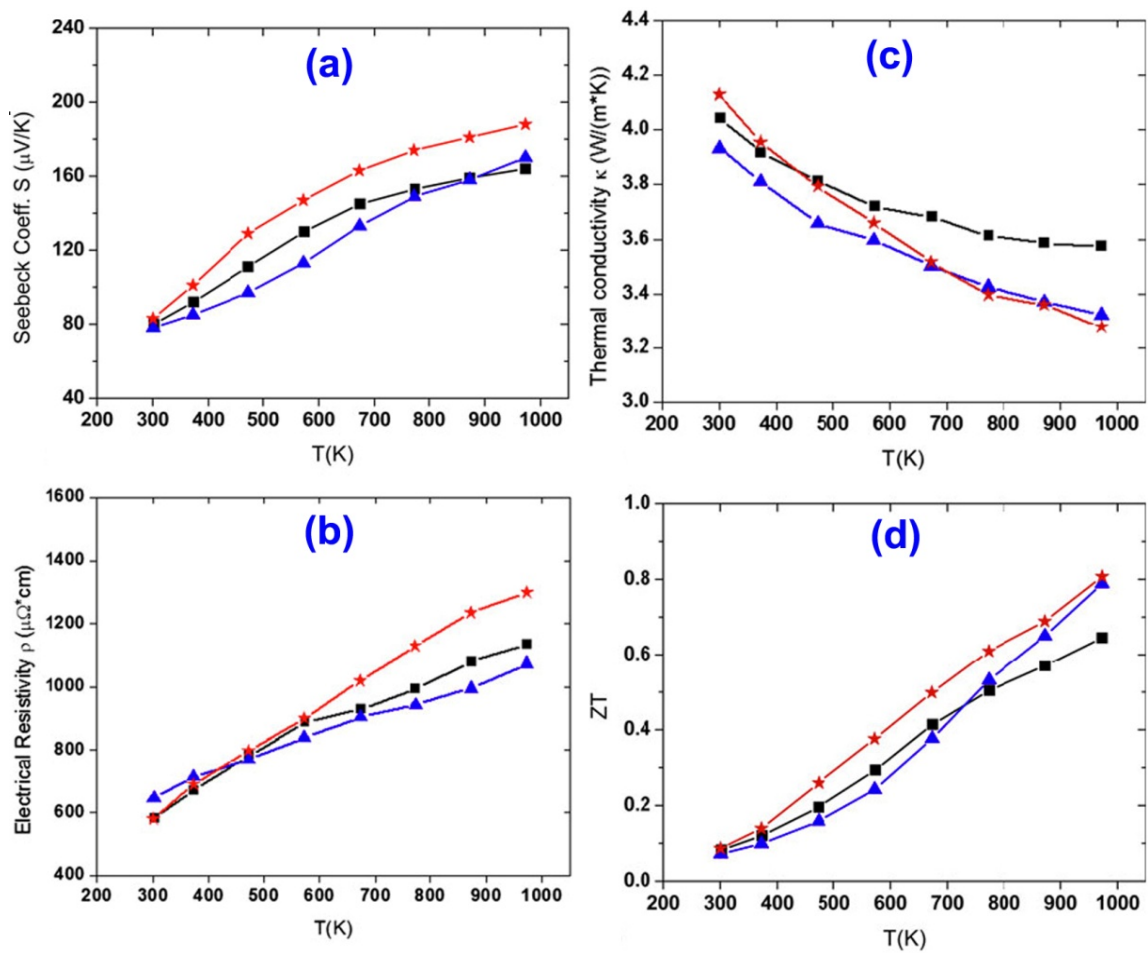

### 3.1.2. *In-Situ* Approach-Nanoscale Precipitation

The *in-situ* approach is a manner or method in which nanoscale inclusions form within the homogeneous matrix via nucleation and growth, such as metastable phase decomposition [116–119] and nanoscale precipitation [120–125].

In the MCoSb (M = Ti, Zr, or Hf) system, when indium (In) was added into the mixed starting materials during the induction melting, indium (In) can react with antimony (Sb) and thus results in the formation of nanoscale InSb inclusions [79]. By the *in-situ* formation of InSb nanoinclusions in the (Ti,Zr,Hf)(Co,Ni)Sb matrix, Xie *et al.* have experimentally shown that the Seebeck coefficient, electrical conductivity, and thermal conductivity can be tuned *somewhat individually*, albeit in a small portion of the overall parameter space. The (Ti,Zr,Hf)(Co,Ni)Sb-InSb nanocomposites have been prepared by high-frequency induction melting method combined with a subsequent SPS process. InSb can be formed *in-situ* during the induction melting, and mainly distribute around the grain boundaries of the HH matrix, as shown in Figure 10. As presented in Figure 10b, the size of InSb nanoinclusions are very uniform and around 10–30 nm in the HH nanocomposites with 1 at.% InSb. However, the typical size of the InSb nanoinclusions can rapidly grow from 10–30 nm for HH nanocomposites with 1 at.% InSb to 200–300 nm for HH nanocomposites with 7 at.% InSb, as the InSb concentration increases. Such nano-scale InSb inclusions "inject" electrons that can increase the carrier concentration and thus the electrical conductivity of (Ti,Zr,Hf)(Co,Ni)Sb-InSb nanocomposites (Figure 11a), and also InSb-(Ti,Zr,Hf)(Co,Ni)Sb interfaces act as potential energy barriers in order to scatter low energy electron resulting in an enhancement of the Seebeck coefficient (Figure 11b). Meanwhile, the phonon scattering is enhanced at the matrix-nanoinclusion interfaces, resulting in a reduction in lattice thermal conductivity (Figure11c). Due to the combination of "electron-filtering", "electron-injection" and "boundary phonon scattering" mechanisms, the otherwise inter-related Seebeck coefficient, electrical conductivity, and thermal conductivity are significantly decoupled. As a result, the *ZT* of the HH nanocomposites with 1 at.% InSb has attained an improvement of ~160% as compared with the HH matrix sample (Figure 11d). Such an *in-situ* nanostructure formation approach has successfully been applied to the p-type Ti(CoFe)Sb system, and the effect of the InSb nanoinclusions on the TE transport properties of Ti(CoFe)Sb-InSb nanocomposites is almost the same as its effect in the *n*-type (Ti,Zr,Hf)(Co,Ni)Sb-InSb nanocomposites [unpublished]. Although the highest *ZT* of both n and p-type HH-InSb nanocomposites increases by 160%–450% as compared with that of HH matrix, the as attained *ZT* value is still lower than that of the benchmark *ZT* ~1 for practical TE materials. This is due to the low *ZT* in the starting HH matrix. *In-situ* formation of the InSb nanoinclusions in a HH matrix with a higher *ZT* will most likely achieve an even higher *ZT* for HH-InSb nanocomposites.

**Figure 10.** Microstructures of nanoinclusions for (Ti,Zr,Hf)(Co,Ni)Sb nanocomposites with 1 at.% (**a** and **b**) and 7 at.% (**c**–**e**) InSb nanoinclusions. Reproduced with permission from Reference [79]. Copyright 2010, Elsevier.

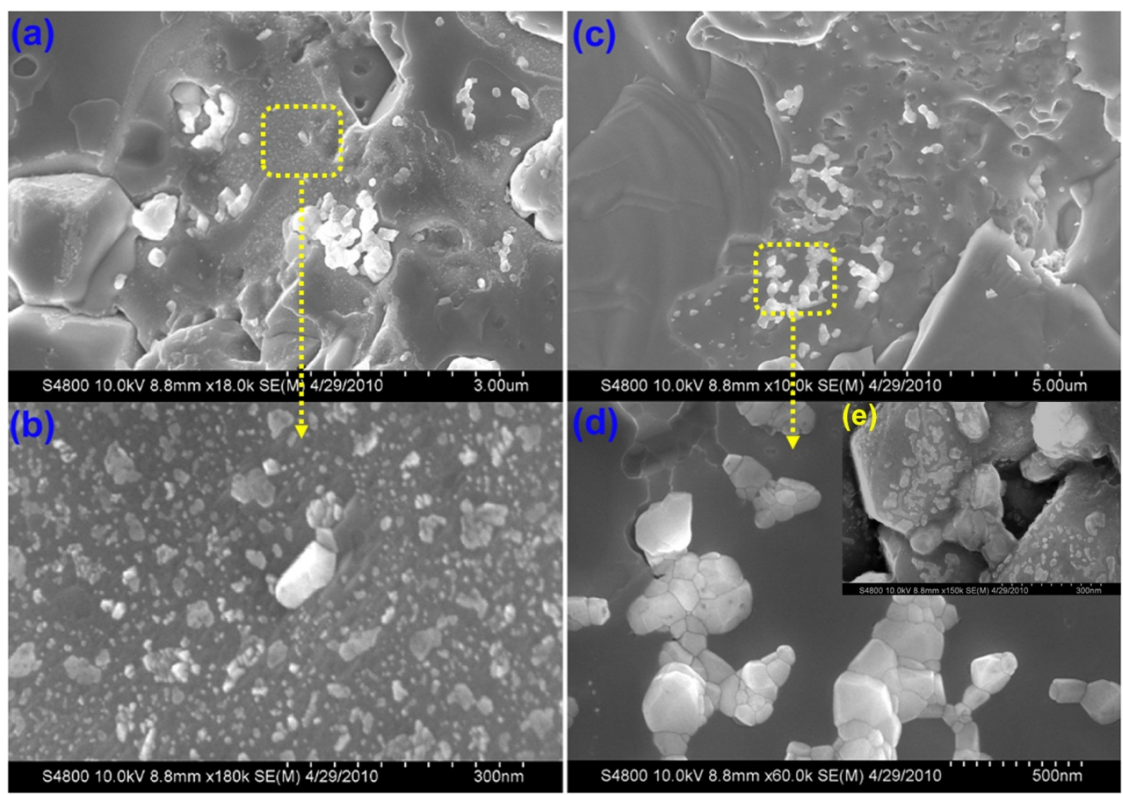

**Figure 11.** TE transport properties of the (Ti,Zr,Hf)(Co,Ni)Sb-x at.% InSb nanocomposites ("HH" represents the formula (Ti,Zr,Hf)(Co,Ni)Sb, and $x$ = 0, 1, 3, and 7.): (**a**) the Seebeck coefficient; (**b**) electrical resistivity; (**c**) thermal conductivity; and (**d**) dimensionless figure of merit *ZT* values. Reproduced with permission from Reference [79]. Copyright 2010, Elsevier.

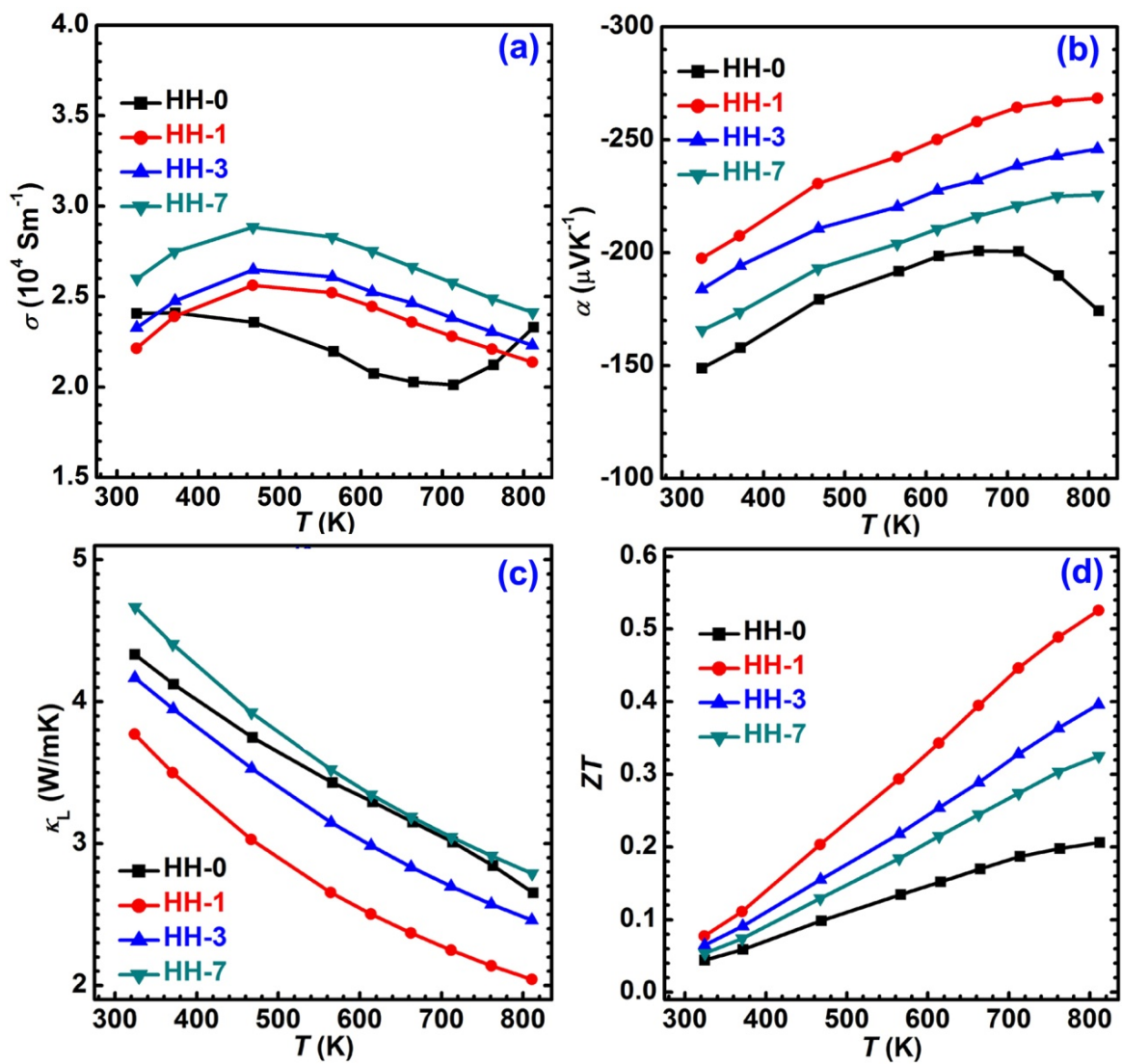

More recently, Poudeu proposed an "Atomic-Scale Structural Engineering of Thermoelectric" (ASSET) approach to *in-situ* form "half Huesler-full Heusler" (HH-FH) nanocomposites [80,106]. In their ASSET approach, an extra 2 atom% elemental Ni was added into as prepared $Zr_{0.25}Hf_{0.75}NiSn$, and the mixed Ni and $Zr_{0.25}Hf_{0.75}NiSn$ was sealed into a silica tube under residual pressure of ~$10^{-4}$ Torr for heat treatment. The full Heusler phase can be in-situ formed via following reaction:

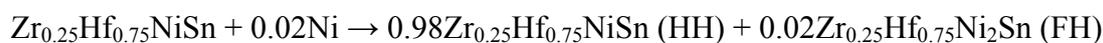

$$Zr_{0.25}Hf_{0.75}NiSn + 0.02Ni \rightarrow 0.98Zr_{0.25}Hf_{0.75}NiSn \text{ (HH)} + 0.02Zr_{0.25}Hf_{0.75}Ni_2Sn \text{ (FH)}$$

Due to the immiscibility of the HH phase $Zr_{0.25}Hf_{0.75}NiSn$ and FH phase $Zr_{0.25}Hf_{0.75}Ni_2Sn$, and the similarity of their crystal structures along with a small lattice mismatch, then $Zr_{0.25}Hf_{0.75}Ni_2Sn$ as a nanoscale inclusion in the HH matrix can be formed *in-situ*. Figure 12 shows the TEM images of $Zr_{0.25}Hf_{0.75}NiSn(1-x)$-$Zr_{0.25}Hf_{0.75}Ni_2Sn(x)$ [HH(1−x)-FH(x)] containing 2 and 5 mol% FH inclusions.

TEM images shown in Figure 12 indicated that spherically shaped nanoscale precipitates are well dispersed in the HH matrix. It worth pointing out that the HH and FH form a coherent phase interface between each other. Chai *et al.* also observed the same situation in the TiNiSn system, and they explained the formation mechanism of the FH as that of vacancies of the Ni atoms move in the TiNiSn lattice and then form a TiNiSn-TiNi$_2$Sn coherent boundary, as shown in Figure 13 [120]. Such small size nanoprecipitates with coherent boundaries indeed significantly influence the TE transport properties of the HH-FH nanocomposites in a way of decoupling the interrelated electrical conductivity, the Seebeck coefficient and thermal conductivity. The temperature dependence of the thermal and electrical properties of HH(1−*x*)-FH(*x*) bulk nanocomposites (*x* = 0.02, 0.05) are shown in Figure 14. In order to distinguish the effect of nanoprecipitates on the TE transport properties, the TE properties of a Zr$_{0.25}$Hf$_{0.75}$NiSn matrix and Sb-doped and a nanostructure-free Zr$_{0.25}$Hf$_{0.75}$NiSn$_{0.975}$Sb$_{0.025}$ bulk half-Heusler alloy are also included for comparison. As shown in Figure 14a, the large increase in the Seebeck coefficient results from the filtering of low energy electrons at the nanometer scale HH/FH phase boundaries leading to a decrease in the carrier concentration around room temperature presented in Figure 14c. The mechanism of the energy filtering effect induced by coherent boundaries of the HH/FH composite is illustrated in Figure 15. This reduction in carrier concentration is compensated by a large increase in the carrier mobility, which in turn minimizes the loss in the electrical conductivity, as shown in Figure 14b. At high temperatures, the electrical conductivity significantly increases due to a quite notable increase in the carrier concentration and slow decrease in carrier mobility. This simultaneous increase in the Seebeck coefficient and electrical conductivity at high temperatures leads to an impressive enhancement of the power factors of the HH-FH nanocomposites (Figure 14e). Furthermore, the lattice thermal conductivity of HH-FH nanocomposites is much lower than that of HH matrix without any nanostructure. This has been attributed to the ability of multiple coherent nanometer scale HH/FH phase boundaries within the nanocomposites to effectively scatter mid-frequency phonons in addition to the typical scattering of mid-to-long wavelength phonons at grain boundaries as well as the scattering of short wavelength phonons by point defects. Although, to date, we have not found any experimental results related to the application of the ASSET strategy on p-type HH nanocomposites, we believe that it is highly recommended to try such an approach.

**Figure 12.** TEM images of spark plasma sintered pellets of HH(1−x)-FH(x) bulk nanocomposites containing 2 mol% (**A**) and 5 mol% (**B**) FH inclusions. The spherical shape of the precipitates suggests their nucleation and isotropic growth as well as their endotaxial insertion within the HH matrix. In addition to spherical precipitates, 2–8 nm thick and up to 30 nm long lamellar structures are also observed in panel B. High magnification image of one of the lamellar fine features revealed details of the HH-FH phase boundary structure (**C**). The FH and HH phases form a tilted coherent interface with tilt angle of approximately θ = 30°. Fast Fourier Transform (FFT) diffractograms of both HH and FH regions showed that the FH inclusions with unit cell parameter a = 6.29(3) Å is oriented along the [1 -1 -2] zone axis (**D**), whereas the HH matrix with unit cell parameter a = 6.06(3) Å is oriented along [-1 2 1] (Figure 1E). Reproduced with permission from Reference [80]. Copyright 2011, American Chemical Society.

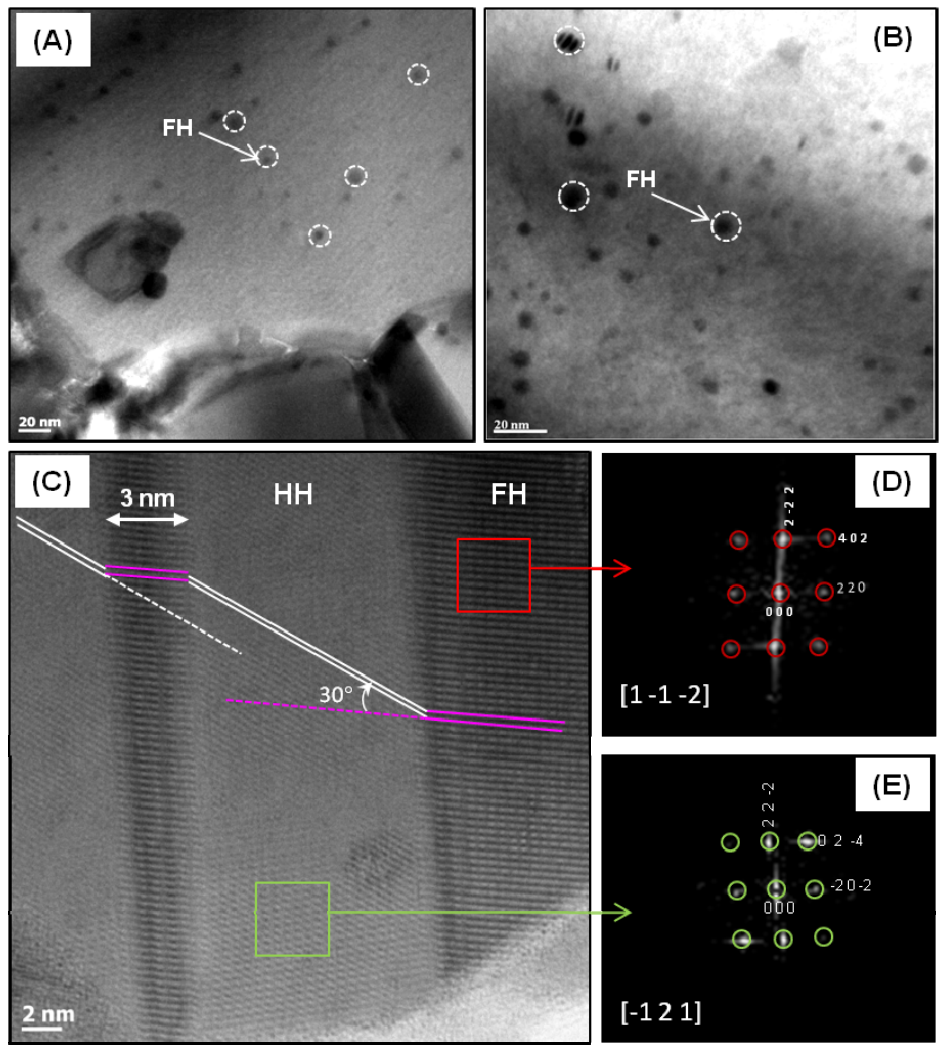

**Figure 13.** Schematic diagram illustrating the formation of a Ni rich cluster through diffusion of Ni atoms into vacant sites (see arrows). The thick solid line at the bottom left hand side indicates a unit cell with HH structure, and dashed lines indicate $[111]_{HH}$ lattice planes. The diagram is viewed along $[10\bar{1}]_{HH}$. Reproduced with permission from Reference [120]. Copyright 2012, American Institute of Physics.

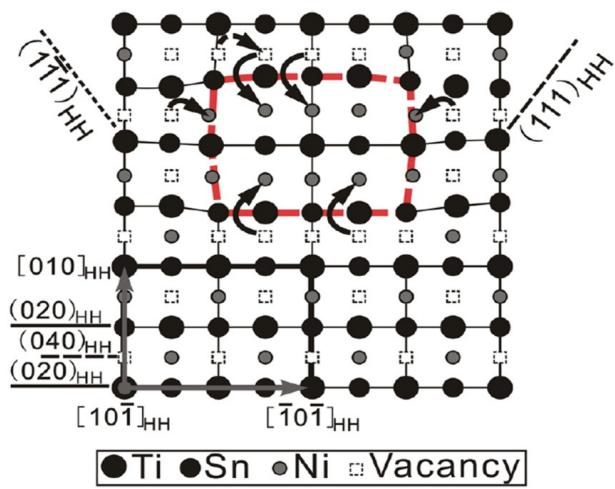

**Figure 14.** Temperature dependence of the thermal and electrical properties of HH(1−*x*)-FH(*x*) bulk nanocomposites (*x* = 0.02, 0.05) compared to that of Zr$_{0.25}$Hf$_{0.75}$NiSn matrix and Sb-doped and nanostructure-free Zr$_{0.25}$Hf$_{0.75}$NiSn$_{0.975}$Sb$_{0.025}$ bulk half-Heusler alloy: (**a**) Seebeck coefficient; (**b**) electrical conductivity; (**c**) carrier concentration; (**d**) carrier mobility; (**e**) power factor; (**f**) lattice thermal conductivity. Reproduced with permission from Reference [80]. Copyright 2011, American Chemical Society.

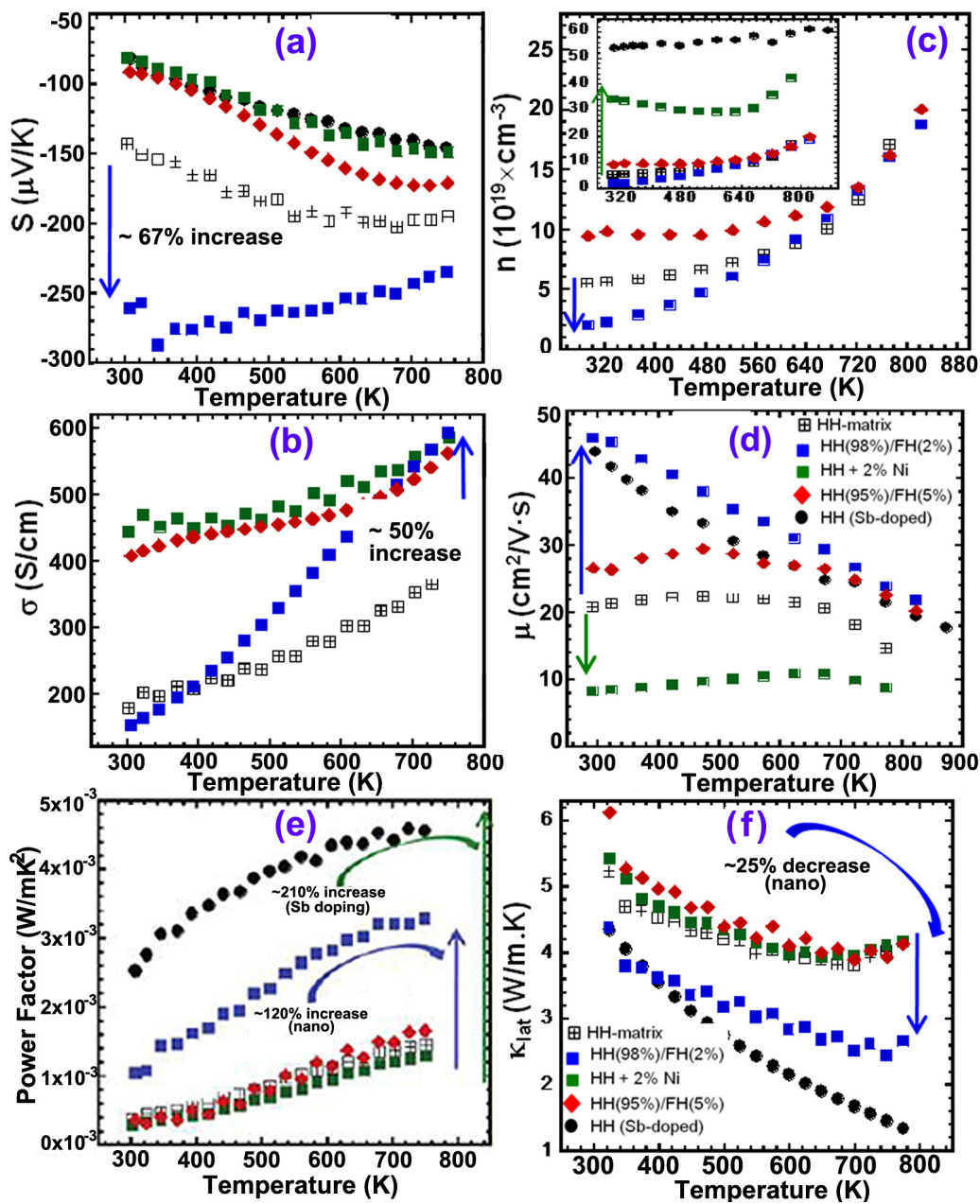

**Figure 15.** Schematic illustration of examples of atomic scale structural engineering leading to large enhancements of the thermoelectric performance of bulk half-Heusler matrices. (**A**) Incoherent interface at grain boundaries (BG); (**B**) coherent phase boundaries (PB) between matrix and inclusion phases with similar structure, which allow close crystallographic registry between both regions; (**C**) nanometer scale heterojunction between bulk HH matrix and nanometer scale FH inclusion highlighting the proposed low energy electron filtering mechanism. The conduction band offset, ΔE, is a critical parameter for a simultaneous large increase in both the Seebeck coefficient and the electrical conductivity through filtering of low energy electrons. Cyan and orange colors represent distributions of low and high energy electrons at temperatures T1 and T2 (T2 > T1). Reproduced with permission from Reference [80]. Copyright 2011, American Chemical Society.

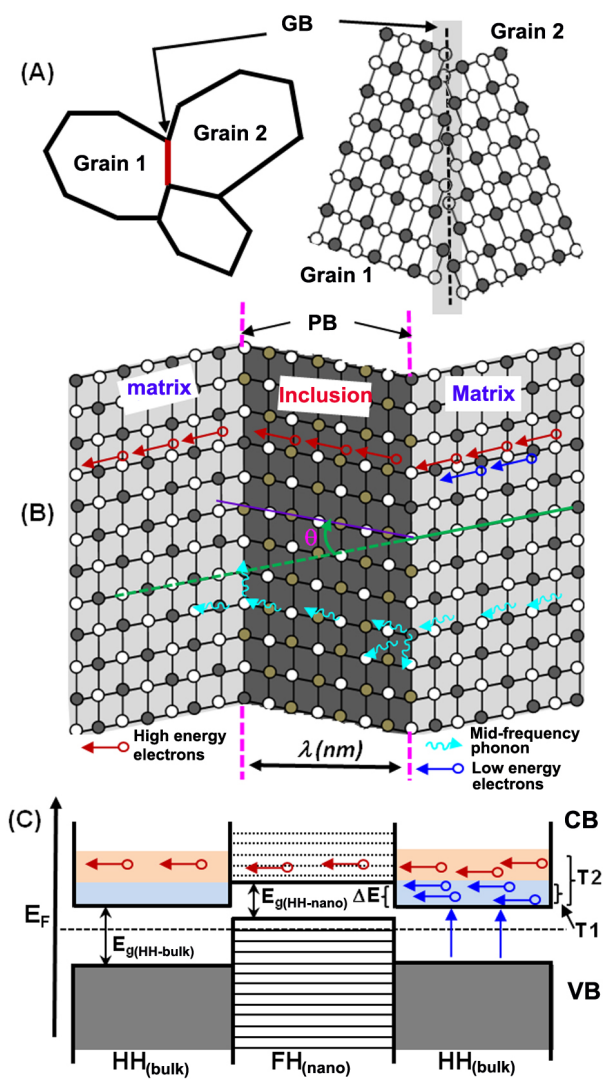

*3.2. Nanoscale HH Matrix with Nanoinclusions*

Usually a nanoscale HH host matrix with nanoinclusions are prepared by a two-step procedure: the HH ingot could be first prepared by arc melting, induction melting, or levitation melting, and then the grain size of ingots are reduced by some further processing in order to form a nanoscale HH matrix. Recently, melt spinning (MS) [126–131] and ball milling (BM) [132–135] techniques were widely used to prepare high performance TE nanocomposites. Both MS and BM are very traditional techniques to prepare nanostructured materials. MS combined with spark plasma sintering (MS-SPS) was adopted by Zhu *et al.* [69,74,76,113] to prepare (Zr,Hf)NiSn based HH nanocomposites. The typical microstructures of the melt spun (Zr,Hf)NiSn based HH nanocomposites are presented in Figure 16. By the MS process, the grain size of HH matrix was reduced to nanoscale dimensions, and also some fine Ni rich FH nanoinclusions were formed after the MS process. Such nano-engineering processes generate numerous grain boundaries that can scatter phonons, resulting in a reduction in lattice thermal conductivity (Figure 17a). Furthermore, the presence of Ni rich FH nanoinclusions causes an enhancement in the electrical conductivity. Despite the simultaneous increase in the electrical conductivity and the reduction in the lattice thermal conductivity, no significant enhancement in the *ZT* was obtained for these MS alloys with the submicron matrix and embedded nanoinclusions due to the concomitantly decreased Seebeck coefficient and increased electronic thermal conductivity.

**Figure 16.** SEM images of the melt spun thin ribbons (**a**) and the fracture surfaces of the MS-SPSed (**b**) $Hf_{0.6}Zr_{0.4}NiSn_{0.98}Sb_{0.02}$. TEM images of the MS $Hf_{0.6}Zr_{0.4}NiSn_{0.98}Sb_{0.02}$ bulk sample, showing the submicron grains (**c**) and embedded in-situ nanophases (**d**). The inset in (**d**) is the high resolution TEM image of a nanophase samples. (**a**) and (**b**) reproduced with permission from Reference [74]. Copyright 2010, Springer. (**c**) and (**d**) reproduced with permission from Reference [69]. Copyright 2012, American Institute of Physics.

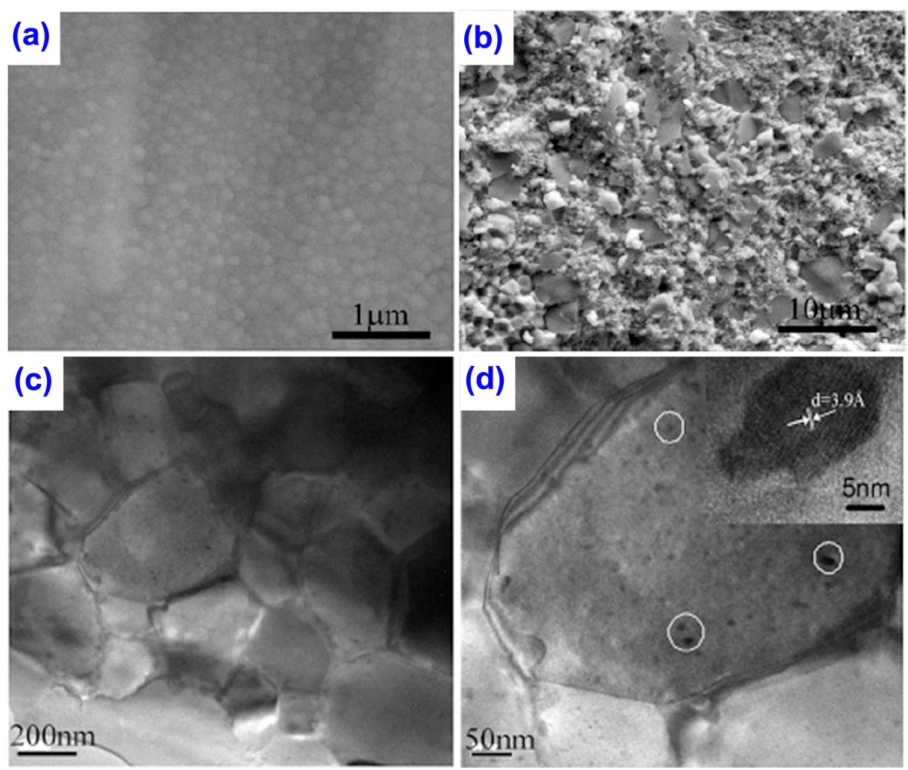

**Figure 17.** Temperature dependence of the thermal conductivity and *ZT* values of levitation melting (LM) and melt spinning (MS) prepared (Hf,Zr)NiSn based HH compounds. The data is adopted and calculated from Reference [74].

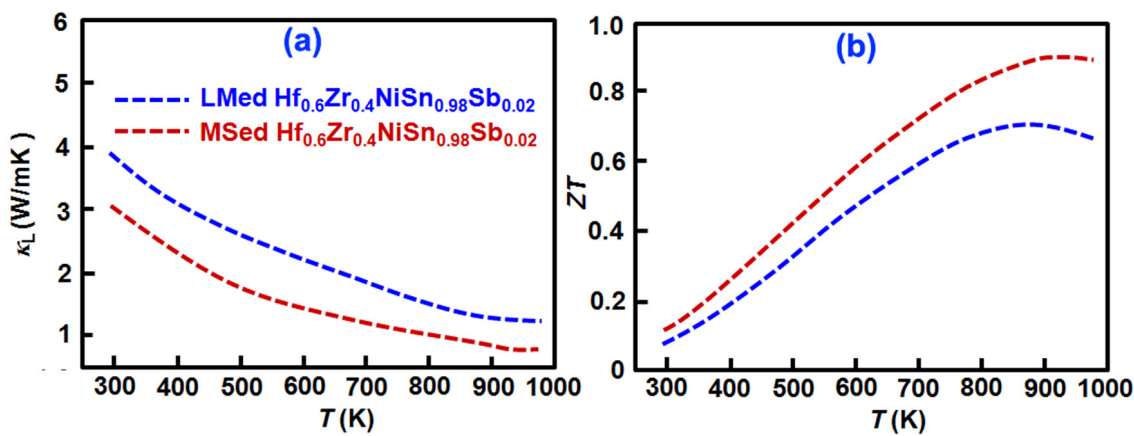

Compared with MS-SPS process, the BM process combined with direct current hot pressing (BM-HP) achieves more success in enhancing *ZT* of HH nanocomposites [72,73]. Using arc-melted HH ingots as starting materials, Ren *et al.* adopted a BM-HP process in order to prepare high TE performance *n* and *p*-type HH nanocomposites. The TEM images (Figure 18a) show that the grain size of the as-prepared *n*-type HH nanocomposite is in the range of around 200–300 nm, some precipitates or aggregates in the matrix (Figure 18c), and the discontinuous heavily distorted crystal lattice, marked by arrows (Figure 18d) exist in the BM-HPed HH nanocomposites. The small grains, precipitates, and lattice distortions are very desirable for lower thermal conductivity due to a possible increase in phonon scattering. The thermal conductivity of BM-HPed HH nanocomposite is much lower than that of the ingot material. However, it worth noting that the electrical conductivity of the *n*-type BM-HPed HH nanocomposite (Figure 19a) significantly decreases by 50% as compared to that of the ingot material at room temperature. The full set of the thermoelectric properties are shown for the *n*-type BM-HPed HH nanocomposite material in (Figure 19a–f). It is not surprising that such small grains, precipitates, and lattice distortions also scatter the charge carriers and thus significantly lower the electrical conductivity. Sharp *et al.* [136] estimates the phonon mean free path as well as electron mean free path for a ZrNiSn compound, and the results show that if the grain size of ZrNiSn is around 200 nm, then nanostructures with such a specific size can more effective at scattering phonons rather than scattering electrons, as shown in Figure 20. However, if the size decreases to tens of nanometer, the gained effect of the reduction in the phonon mean free path will be much less than that of the reduction in electron mean free path. The same situation is also that which may have occurred in the BM-HPed p-type HH nanocomposite. So far, the BM-HPed HH nanocomposite have achieved the highest *ZT* ≈ 1.0 in n-type Hf$_{0.75}$Zr$_{0.25}$NiSn$_{0.99}$Sb$_{0.01}$ nanocomposites [72], and *ZT* =1.0 in *p*-type HfZrCoSbSn [73] from the *ZT* = 0.8 and *ZT* = 0.6 of their ingot host materials, respectively.

**Figure 18.** (**a**) Low- and (**b**–**d**) high-magnification TEM images of a BM-HPed $Hf_{0.75}Zr_{0.25}NiSn_{0.99}Sb_{0.01}$ nanocomposite. The inset in (**b**) shows the crystalline nature of grain 1 with a rotation. The inset in (**d**) shows the grains as good crystalline structures. Reproduced with permission from Reference [72]. Copyright 2011, John Wiley and Sons.

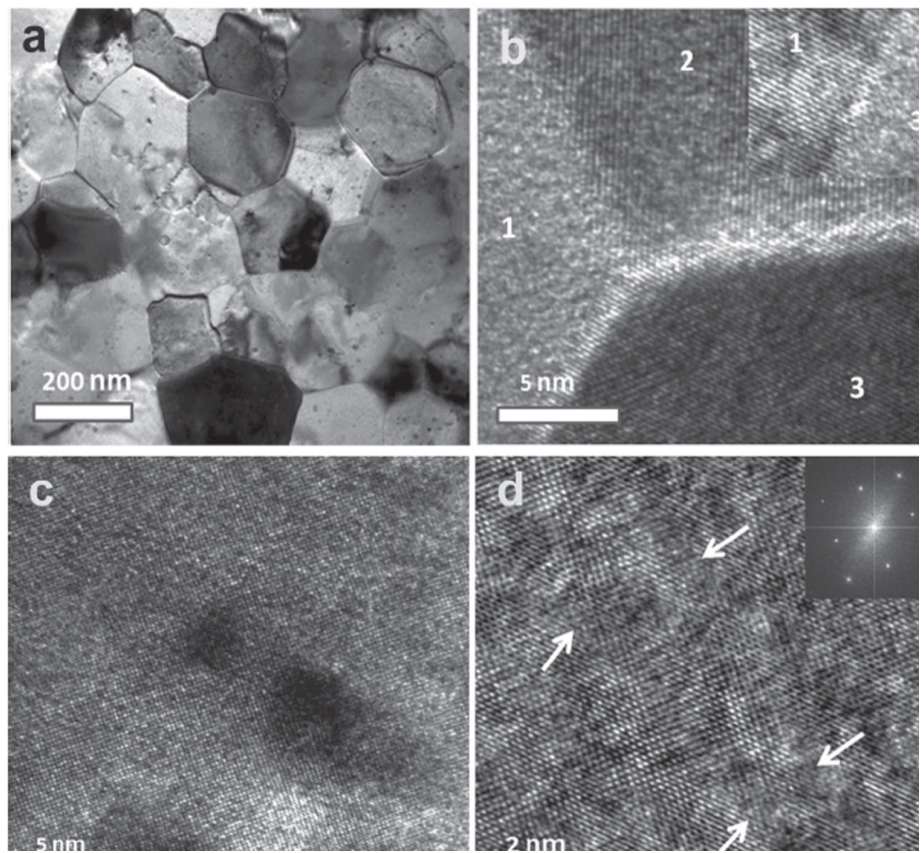

**Figure 19.** (**a**) Temperature-dependent electrical conductivity, (**b**) Seebeck coefficient, (**c**) power factor, (**d**) total thermal conductivity, (**e**) lattice thermal conductivity, and (**f**) *ZT* values of three nanostructured $Hf_{0.75}Zr_{0.25}NiSn_{0.99}Sb_{0.01}$ samples, and the sample annealed at 800 °C for 12 h in air; the line is for viewing guidance only, for comparison with the ingot sample (open circles) which matches the previously reported best n-type half-Heusler composition. Reproduced with permission from Reference [72]. Copyright 2011, John Wiley and Sons.

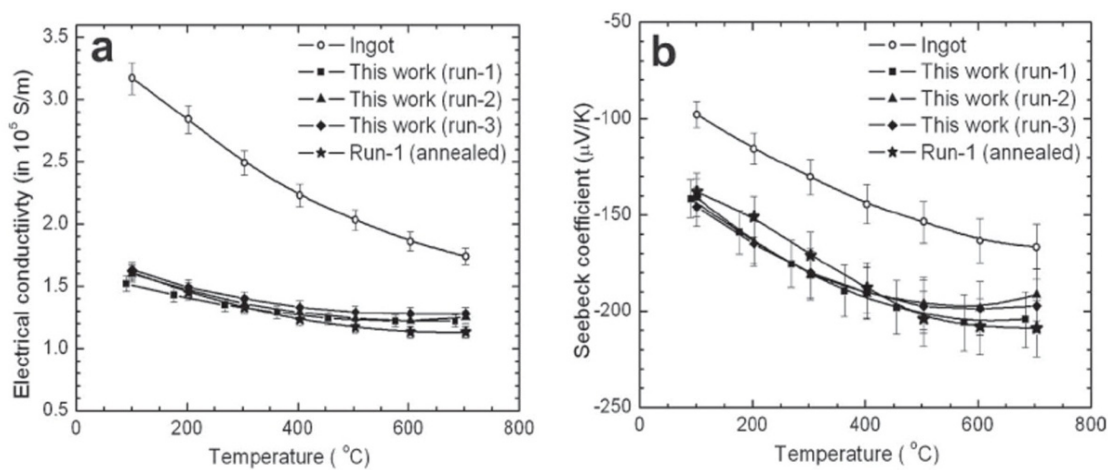

Figure 19. *Cont.*

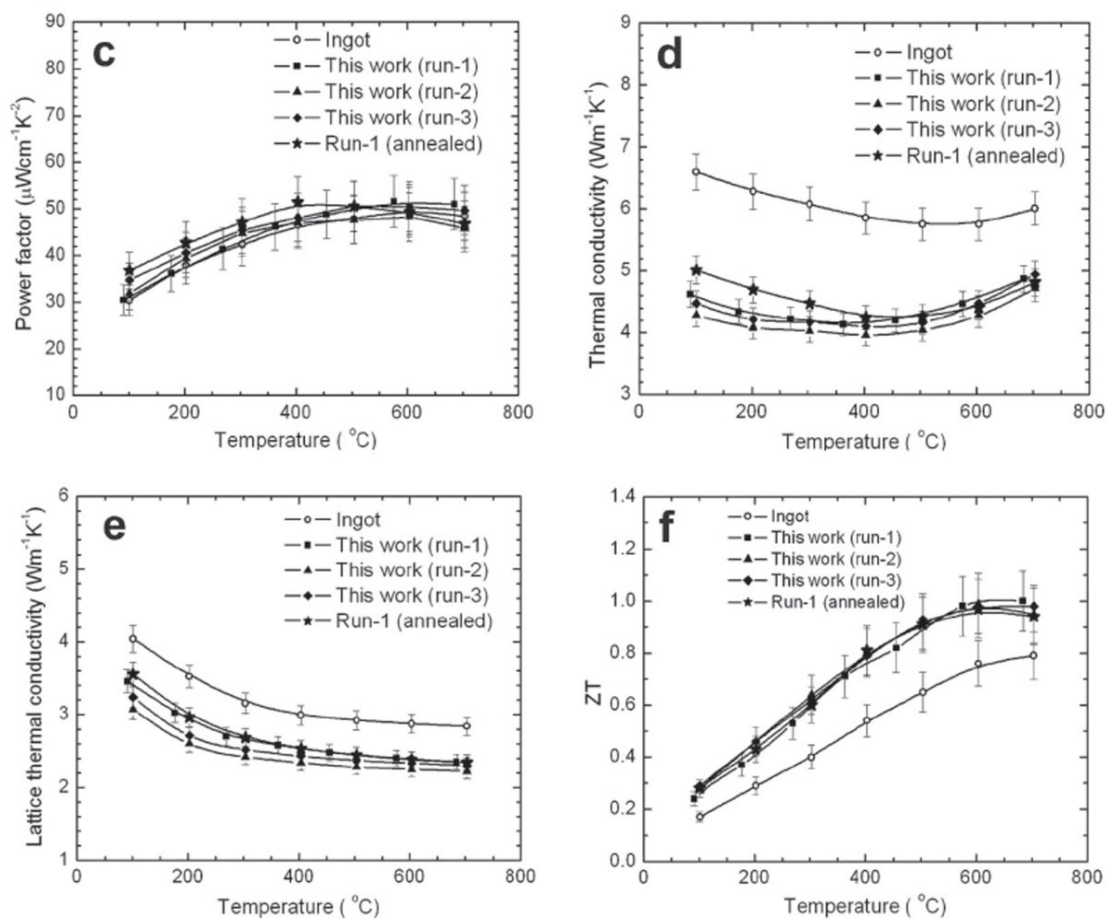

**Figure 20.** Boundary scattering in $Zr_{0.5}Hf_{0.5}NiSn$. Reproduced with permission from Reference [136]. Copyright 2001, John Wiley and Sons.

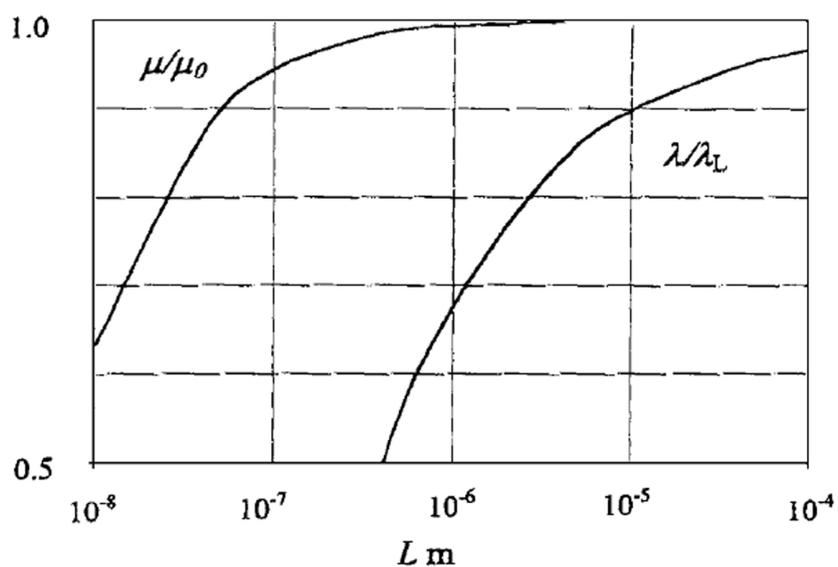

## 4. The Future and Challenge of Nanostructuring in HH Compounds

Figure 21 shows the overall trend in the development of the *ZT* (TE performance) of HH compounds over last two decades. First, doping/substitution can optimize the carrier concentration to achieve the highest power factor as well as *ZT* values. Furthermore, the *ZT* values of doped HH compounds can be further enhanced by the nanostructuring approach due to boundary scattering effect and energy filtering effect. Now one should ask the question: what is the next step to enhance *ZT*?

**Figure 21.** The *ZT* values of HH compounds are significantly enhanced by doping and nanostructuring in last two decades: (**a**) *n*-type HH; (**b**) *p*-type HH.

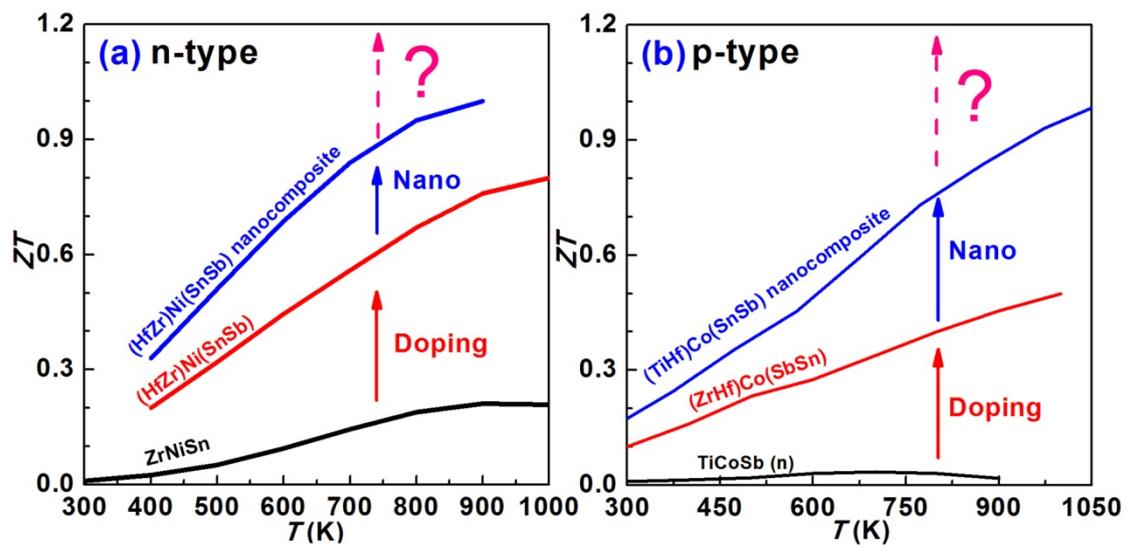

One such question; is there still some ability to further reduce the lattice thermal conductivity by nanostructuring? In order to investigate just how much more further reduction of the lattice thermal conductivity are possible, then we need to estimate the minimum lattice thermal conductivity, $\kappa_{min}$, for the specific materials. For example, $\kappa_{min}$ is estimated for the ZrNiSn compound by applying a model developed by Cahill *et al.* [137]:

$$\kappa_{min} = (\frac{\pi}{6})^{1/3} k_B n_a^{2/3} \sum_i v_i (\frac{T}{\theta_i})^2 \int_0^{\theta_i/T} \frac{x^3 e^x}{(e^x-1)^2} dx \qquad (3)$$

where the summation is over the three polarization modes and $k_B$ the Boltzmann constant. The cut-off frequency (in unit of Kelvin) is $\theta_i = v_i(\hbar/k_B)(6\pi^2 n_a)^{1/3}$, where $n_a$ is the number density of atoms, $\hbar$ the reduced Planck constant, $v_i$ the sound velocity for each polarization modes. The $\kappa_{min}$ of ZrNiSn, calculated by Zhu *et al.* [138], is shown in Figure 22, and thermal conductivity of ZrNiSn based nanocomposites are included for comparison. From this plot (Figure 22) it is apparent that there is still some opportunity of further significant reduction of $\kappa_L$ between the $\kappa_{min}$ of ZrNiSn and the lowest $\kappa_L$ of ZrNiSn based nanocomposites in the higher temperature range. Immediately, one might propose that nanostructures with even smaller sizes could further reduce the lattice thermal conductivity for a ZrNiSn based nanocomposite. However, the estimated phonon mean free path of HH-x nanocomposite is on the order of ~1 nm at high temperature [unpublished], which is comparable to the lattice parameter of the HH compound, so this approach is not likely to be effective. And, we should keep in mind that if the size of the HH matrix decreases to ~1 nm the effect of the small grains will also significantly decrease the carrier mobility through enhanced electron scattering [136]. Therefore, just blindly pursuing even smaller nanostructure is not likely to lead to the desired effect of enhancing the *ZT* but would most likely lead to a decrease in the *ZT*.

**Figure 22.** The lattice thermal conductivity of MSed $Zr_{0.4}Hf_{0.6}NiSn_{0.98}Sb_{0.02}$ nanocomposite and $k_{min}$ of ZrNiSn estimated by applying a model developed by Cahill *et al.* [137]. The data is adopted from Reference [138].

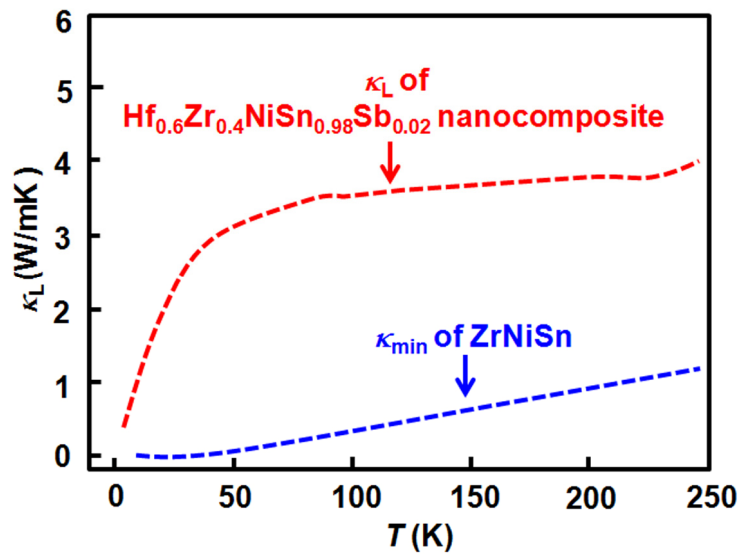

In addition, besides reducing the lattice thermal conductivity, one should also be more attentive of ways to further improve the power factor by nanostructuring, especially enhancing the Seebeck coefficient. The above discussion mentioned that the special potential energy barrier generated by nanoinclusions can selectively scatter low energy carriers and thus result in a significant improvement in the Seebeck coefficient. At the same time, if the nanoinclusions can donate high mobility carriers into the matrix to compensate for the carrier loss in the energy filtering process, the enhancement in Seebeck coefficient should not significantly reduce the electrical conductivity. Such specific nanoinclusions can induce the energy filtering effect, the carrier injection effect and the boundary scattering effect to simultaneously improve the Seebeck coefficient, increase electrical conductivity, and decrease lattice thermal conductivity. Such cases are observed in micro-scale HH matrix with nanoinclusions [79,80], but not in nanoscale HH matrix with nanoinclusions. However, the first significant challenge is how to choose the correct secondary phase as the best nanoinclusions. According to previous experimental results, it seems that a metallic phase (e.g., full Heusler phase $Hf_{0.75}Zr_{0.25}Ni_2Sn$) or a semiconductor with highly carrier mobility (e.g., InSb) might be able to fulfill such requirements. Nevertheless, the theoretical calculations and the theoretical ability to predict the behavior of such materials is still quite lagging the experimental work in this field. Therefore, more theoretical calculations and modeling in order to produce predictive directions are highly desirable in order to better direct the most promising routes or directions to obtain the most favorable TE nanoinclusions.

Moreover, the above reviewed nanocomposites have two common characteristics: (1) nanoinclusions are randomly distributed, and (2) their sizes are to a large extent uncontrollable. If the nanoinclusions could uniformly distribute in the matrix, and also the size can be more precisely controlled, it would be much more straightforward to investigate how the size and the distribution of nanoinclusions affect the TE transport properties. Then the most favorable distribution of size, position and distribution for nanoinclusions for the best TE performance can be found. Although Chen points out that the surface per volume (S/V) ratio is more important than the periodicity distribution of the nanostructure for the reduction in the lattice thermal conductivity, it would be even more desirable if the uniform distribution of the nanostructures can even enhance the electrical conductivity. It is not expected that nanostructures result in strong carrier scattering, and also defects induced by nanostructures can dramatically change the carrier concentration. Recently Liu *et al.* [139] points out that ordered nanocomposites with various well-organized (distributed) nanostructures may reconstruct a carrier transport channel, as to depress lattice thermal conductivity without significantly affecting the electrical conductivity. The second open challenge for theorists and experimentalists is how the size and distribution of nanoinclusions can be more precisely controlled within a certain TE matrix and this has been explicitly been shown here for the case on half Huesler alloys.

## 5. Conclusions and Outlook

One of the newest directions in TE materials research over the past two decades is the concept of *nanostructure-engineering*. Certainly *nanostructuring or nanocomposites* is the new paradigm of TE materials research. In the HH nanocomposite materials, the TE performance can be significantly enhanced by decreasing the lattice thermal conductivity as well as enhancing the power factor. Nanoinclusions that can induce both a boundary scattering effect and energy filtering effect can effectively decouple the interrelated thermal and electrical transport properties. After numerous efforts in 20 years (or so), the highest *ZT* values of HH compound is pushing to 1–1.2 in HH nanocomposites from 0.1–0.2 in undoped HH ingot. At this date, it is unknown how far nanostructuring can further achieve progress in these materials. Continuing to follow the concept of nanostructuring is a positive direction, but exploring several other scientific and engineering directions for further enhancing *ZT* of HH materials should be pursued as well as achieving device applications of these HH nanocomposites.

(1) Two new physical ideas related to band structure engineering were introduced in TE research field, such as distortion of the electronic density of states (resonant energy level) in Tl doped PbTe [140], and convergence of electronic bands in Na-doped $PbTe_{1-x}Se_x$ [141]. The idea of introducing a resonant level in HH compound was achieved in V doped $Hf_{0.75}Zr_{0.25}NiSn$ HH compound [142]. Even though, the maximum *PF* of V doped $Hf_{0.75}Zr_{0.25}NiSn$ does not result in the best working temperature range (above 700 K), the band structure engineering strategy, such as distortion of the electronic density of states, is highly recommended in the future work.

(2) One very positive aspect of the HH alloys in their bulk form in their very favorable thermal stability. But, several points should be taken into account with regards to potential HH applications. First, the mechanical properties of HH nanocomposites should be more fully understood before designing HH modules; second the thermal stability of nanostructures in HH nanocomposites should be more thoroughly investigated (even though the matrix has been shown to be very stable); and third, the possibility of scale-up production for high TE performance HH nanocomposites should be further investigated and evaluated.

Although to date, the current TE materials cannot be in used in very broad applications due to their low conversion efficiency. However, if the combination of combined mechanisms to enhance ZT in HH alloys, such as *nanostructuring and band structure engineering*, are able to achieve values of the *ZT* to 2 or even higher, improve the mechanical properties and maintain their good thermal stability with the nanostructuring processes, then the broad range of potential applications for waste heat recovery using TE devices based on this next generation of HH alloys will be very promising.

## Acknowledgements

W.J.X. and A.W. thank the support from EMPA and Swisselectric. W.J.X. would also like to thank the Marie Curie COFUND fellowship supported by EU FY7 and EMPA. T.M.T. acknowledges support from a DOE/EPSCoR Implementation Grant (#DE-FG02-04ER-46139), and the SC EPSCoR cost-sharing program.). X.F.T. and Q.J.Z. acknowledge the support received from the International Science & Technology Cooperation Program of China (Grant No. 2011DFB60150), National Basic Research Program of China (Grant No. 2007CB607501), as well as 111 Project (Grant No. B07040).